# Tolman test from z=0.1 to z=5.5: Preliminary results challenge the expanding Universe model

Eric J. Lerner
Lawrenceville Plasma Physics, Inc., USA
elerner@igc.org

**Abstract**

We performed the Tolman surface-brightness test for the expansion of the universe using a large UV dataset of disk galaxies in a wide range of redshifts (from 0.03 to 5.7). We combined data for low-z galaxies from GALEX observations with those for high-z objects from HST UltraDeep Field images. Starting from the data in publicly-available GALEX and UDF catalogs, we created 6 samples of galaxies with observations in a rest-frame band centered at 141 nm and 5 with data from one centered on 225 nm. These bands correspond, respectively, to the FUV and NUV bands of GALEX for objects at z = 0.1. By maintaining the same rest-frame wave-band of all observations we greatly minimized the effects of k-correction and filter transformation. Since SB depends on the absolute magnitude, all galaxy samples were then matched for the absolute magnitude range (-17.7 < M(AB) < -19.0) and for mean absolute magnitude. We performed homogeneous measurements of the magnitude and half-light radius for all the galaxies in the 11 samples, obtaining the median UV surface brightness for each sample.

We compared the data with two models: 1) The LCDM expanding universe model with the widely-accepted evolution of galaxy size $R = R_0\ H(z)^{-1}$ and 2) a simple, Euclidean, non-expanding (ENE) model with the distance given by $d=cz/H_0$. We found that the ENE model was a significantly better fit to the data than the LCDM model with galaxy size evolution. While the LCDM model provides a good fit to the HUDF data alone, there is a 1.2 magnitude difference in the SB predicted from the model for the GALEX data and observations, a difference at least 5 times larger than any statistical error. The ENE provides a good fit to all the data except the two points with z>4, and there are inconsistencies in the GALEX measurements when using this model. Given the importance of any test of the expansion of the universe, we intend to check these preliminary results with additional data.

keywords{Galaxies: general , Cosmology: general, Tolman test}

## 1. Introduction

As Tolman demonstrated (1930), any expanding universe model predicts a decrease in surface brightness (SB) of identical objects by $(z+1)^4$, where z is redshift and where SB is measured in bolometric units (VEGA-magnitudes/arsec$^2$ or erg/sec cm$^2$ arcsec$^2$) . One factor of (z+1) is due to the time-dilation factor (decrease in photons per unit time), one factor is from the decrease in energy carried by each photon, while two additional factors of (z+1) are due to the object being closer to us by a factor of (z+1) at the time the light was emitted and thus having a larger angular size. (If AB magnitudes or flux/Hz units are used, the dimming is by a factor of $(z+1)^3$, while for ST magnitude or flux per nm units, the dimming is by a factor of $(z+1)^5$ . The dimming is the same for all expanding universe geometries, independent of the specific parameters of the cosmological model.

By contrast , in a Euclidean, non-expanding (ENE) universe, where the redshift is not due to expansion, but to some other physical process, the SB decreases as (1+z) in VEGA-magnitudes and is independent of distance in AB magnitudes per $arcsec^2$. In this paper, we use AB magnitudes throughout, so the ENE prediction is that SB is constant. The difference between the two models, $(z+1)^3$ in all units, is independent of the specific cosmological parameters used and is a very large factor at the maximum redshifts now observable—a factor of 216, for example, at a redshift of 5.

So, it should be possible to use the Tolman test to distinguish the dark-energy, cold dark matter (LCDM) expanding and ENE models. There have been a number of efforts to do this in the past. But, as Scarpa et al (2009) describes in detail, these previous attempts have been seriously limited in various ways. Some papers were highly limited in redshift range: Pahre et al (1996) to z<0.4, Lubin and Sandage (2001) to <0.9. Weedman et al (1998) and Hathi et al (2008) both restricted their studies to maximum surface brightness for their samples rather than average surface brightnesses, which prevented a self-consistent test of the ENE model. Despite the limitations of these earlier studies, when their data is re-examined in a consistent manner, Scarpa et al show that in each case the data are entirely consistent with the ENE model.

For a decisive test, better data sets are needed. Fortunately such data sets have become available since 2005 with the distribution of Hubble Ultra Deep Field (HUDF) images and catalogs and the GALEX Medium Imaging Survey (MIS) catalog. Together these data sets make possible a consistent comparison of surface brightness in the Far UV (FUV) and Near UV (NUV) bands across a range of z from close to 0 to almost 6. Based on a preliminary examination of this data, Lerner (2006) demonstrated that again, there was general consistency with the non-expanding expectations. Motivated by this interesting suggestion we have undertaken a detailed analysis of the GALEX and HUDF datasets with new measurements of the all relevant parameters.

The present paper seeks to overcome the limitations of previous work and to provide a comprehensive test of the surface brightness predictions of the two models. To do this, we select galaxies from the HUDF catalog of Coe et al (2006) and perform our own measurements of radius, magnitude and surface brightness in the HUDF B-band (435 nm), V-band (606nm), i-band (775 nm ) and z-band (905 nm) images. We compare galaxy samples based on these images with GALEX galaxies at low redshift. In this way we are able to carry out a consistent comparison of galaxies at the same rest-frame or at-the-galaxy wavelengths, thus minimizing the k-corrections needed when rest-frame wavelengths are different. We also are able to compare galaxies of the same absolute luminosity across a broad range of z, with almost continuous coverage in the range from 0.9 to 5.7.

In Sect. 2 we report on the selection of the objects and describe the construction of the datasets at various redshifts. In Sect. 3 we present our measurements and comparison with the expectations with standard LCDM and ENE models. Finally in Sect. 4 we draw conclusionws and propose new tests.

## 2. The sample selection and datasets

In order to compare SB of galaxies over a significant redshift range (up to z = 5.5) and to avoid strong k-corrections we need to use observation taken in different bands (depending on the redshift) such that the same rest frame region of all the galaxies is compared. At present this requirement is well met only in the UV, combining UV data collected by the GALEX for low z objects with optical observations from HST for high z sources.

GALEX data are used to assess a low z sample of galaxies based on the observations gathered in the two GALEX bands: FUV (155 nm) and NUV (230 nm). We set the average redshift for our low z sample to z=0.1 as the best compromise in order to have a significant number of objects and to use galaxies that are well resolved by GALEX. These data correspond to rest frame observations obtained 141 nm and 209 nm for FUV and NUV filters, respectively.

A number of high z galaxy samples are then selected from the Hubble Ultra Deep Field using the ACS observations obtained in b (435 nm), V(606nm), i (775 nm ) and z (905 nm) bands. The redshift of the high z samples is set such as to minimize the k-corrections and thus for each filter it is centered as close as possible at the same rest frame wavelengths of the two low z GALEX samples. Using the 4 ACS filter given above this approach allow us to define 8 high z samples of galaxies, 4 that match the FUV band and 4 for the NUV.

Another important issue for this study is that since SB is correlated with the absolute magnitude M we must compare galaxies that have on average the same M. All the considered samples were thus matched in order to have the same mean absolute magnitude. We selected an absolute magnitude range of -17.7 to -19 in order to include the most luminous galaxies that are visible at all redshifts, to have adequate sample size and to avoid too large a range in luminosity. As described below, we used the same absolute magnitude range in the tests of each cosmological model, using that model's formula for absolute magnitude. Since the two models' formulae actually agreed quite closely, almost the same samples are used for both tests.

In this way we are able to carry out a consistent comparison of low z galaxies across a broad range of z, with almost continuous coverage in the range from z= 0.9 to z-5.7. In the following sections we give the full descriptions of the selection of objects in the considered samples and the details of the corrections applied.

*The low z GALEX dataset*

The low-z sample of disk galaxies was selected from the GALEX Medium Imaging Survey (MIS) catalog. (http://galex.stsci.edu/GR2/)This includes about 27000 (26722) galaxies that are also spectroscopically classified in the SDSS catalog. In order to eliminate the unresolved galaxies we conservatively selected for measurement only those with stellarity < 0.4, although we also counted the number of unresolved galaxies. A further selection criterion is that all the galaxies included be disk galaxies. This is an important consideration to avoid confusion in the data. If a disk galaxy's exponential disk is too dim to be observed in a given image, but its bright bulge, with higher surface brightness, is observed, the result can be confused with a much smaller elliptical galaxy. As will be shown below, the galaxies that we are focusing on, which are very luminous in the UV, are all disk galaxies in the GALEX sample.

We initially chose a narrow redshift range from 0.095 to 0.1 for the GALEX sample. However, in order to test the effects of resolution on the measurements, we also selected a lower z sample in the FUV extending from z=0.02 to 0.05. The final z=0.1 samples of GALEX disk galaxies consist of 40 objects in the FUV sample and 70 objects in the NUV sample. The z=0.04 FUV sample is 29 objects.

*The high z HUDF dataset*

The high-z samples are selected from the HUDF photometric and morphological catalogs (Coe et al.). These catalogs contain photometric measurements for each galaxy in the b, v, i, z, H and J bands, and morphological measurements in the i-band. Each galaxy has a photometric redshift, estimated by two methods: Bayesian Probability (BPZ) and Maximum Likelihood (BML). Coe et al. report that comparison of BPZ with spectroscopic redshifts in the small sample where they are available indicates that, except for a few outliers, BPZ redshifts are accurate to 0.04. To eliminate outliers, we have chosen to use the difference between the BML and BPZ redshifts as an indicator of the reliability of BPZ redshifts, and have eliminated all those with ABS(z(BPZ)-z(BML)) > 0.5. Also in this case we have eliminated all galaxies with stellarity > 0.4, those with Sersic number> 2.5 and with uncertainties in the Sersic number > 1 (all parameters determined from i-band images).

The redshift ranges of the HUDF samples are chosen to match as much as possible the same rest frame region observed for the low z GALEX samples. The upper and lower limits of each redshift range are set so that the central wavelength of the GALEX filter coincides with the upper and lower cutoffs of the HST filter. However in some cases the upper cutoff was smaller to avoid allowing the Lyman-alpha region to enter into the observed rest frame spectral range.

*The SB versus Absolute Magnitude relation*

In order to compare surface brightness, we must ensure that we are comparing similar objects. Since SB is correlated with the absolute magnitude M we must compare galaxies that have on average the same M.

For a non-expanding model, M can be calculated from the apparent magnitude m and the distance d, where d is in Mpc, $M = m - 35 - 5 \log d$. However, a relationship must be assumed between the observed redshift z and d. For this study, we assume that the relationship d= cz/H0 holds for all z. For a non-expanding model, M can be derived from the apparent magnitude m (in the AB system) using the relation: $M - m = 5 - 5\mathrm{Log}(cz/H_0)$.

This assumption is motivated by two considerations. First, the linear Hubble relationship is observed to hold in the lowest redshift region, where there are the best independent tests of distance, so the simplest hypothesis is to extrapolate this relation for all z. Second, the luminosity vs. redshift formula derived from this simple assumption is extremely close to the luminosity relationship based on the LCDM formula, not deviating from it by more than 0.5 magnitudes up to a z of 5.7.

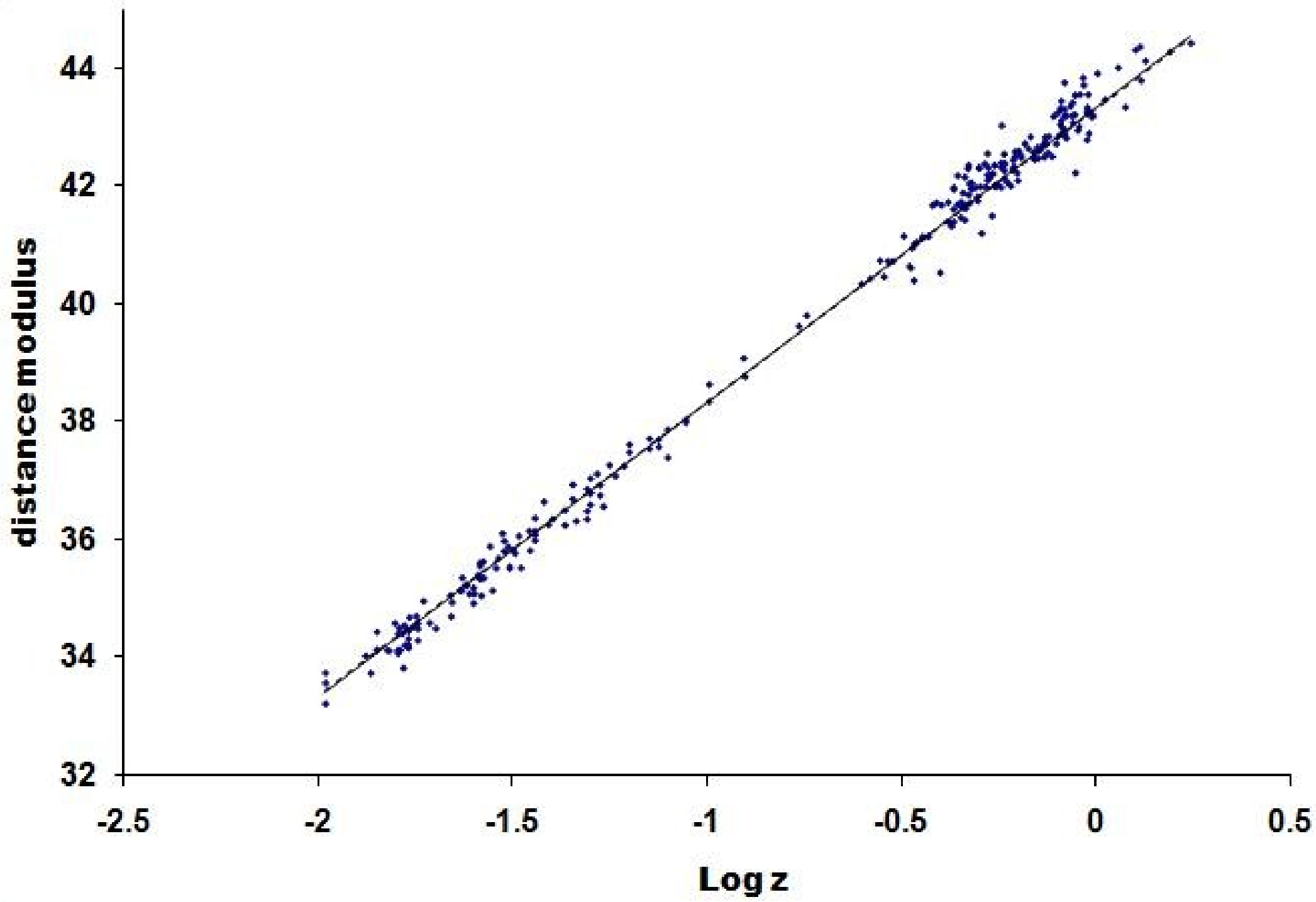

**Figure 1**. Comparison of the distance modulus as a function of the redshift for ENE model (solid line) ($M - m = 5\text{-}5\mathrm{Log}(cz/H_0)$ in AB magnitudes) and the one obtained from the concordance cosmology with Omega_M=0.26 and Omega $\lambda$=0.76 (dashed line). Superposed to the models is the Hubble diagram for supernovae type Ia from the gold sample (Riess04), and the supernovae legacy survey (Astier06). The fits assume the absolute magnitude of supernovae is M= -19.25. The two lines are nearly identical over the whole redshift range.

It is remarkable that this relation gives very similar values (see Figure 1) to those computed using the concordance cosmology. The agreement is better than 0.3 mag over the whole range of redshift up to z =5 and for most of the range, including nearly the entire range covered by supernova observations, it is better than 0.1 mag.

It is important that for both Universe models that we want to compare the relationship between the apparent and absolute magnitude turned out to be almost the same. This allow us to perform the selection of the galaxies based on a given range of absolute magnitudes that is almost independent on the cosmological model assumed. We find that there is almost a complete overlap between the samples chosen using the ENE and LCDM magnitudes, except for the two highest-z samples, and even there most of the samples are the same.

## 3. Data analysis and results

To accurately compare the surface brightness of galaxies, we must use a measurement technique that is consistent for all samples and that is independent of the size of the galaxy images. We could not use the catalog values for GALEX and HUDF, since, first of all, the radii values were derived by different fitting approaches, and second, there were not HUDF catalog values for wavebands other than i-band.

To determine the magnitudes of the galaxies, we took the total flux within a circular aperture. The size of the aperture had to be large enough to accurately measure the light from the largest galaxies, but small enough so as to avoid the contamination from other sources close to the target. Galaxies are excluded if there is a neighbor within 10 pixels of the aperture.

To measure the galaxy radii consistently, we have to avoid several sources of error that would bias the result depending on image size. In galaxies with small image size, the point-spread function (PSF) blurs the central region, making the galaxy appear larger than it is. For galaxies with large images, by contrast, the limit of the aperture cuts off part of their light, so the half-light radius can appear smaller than in reality.

Since we have limited our sample to exponential disk galaxies, a way to avoid these errors is to use the slope of the surface-brightness profile as a measure of radius. The inner 3 pixels are excluded from the fit to avoid the effects of the PSF. However, a simple straight- line fit from the innermost to the outmost part of a galaxy exaggerates the importance of the outermost points, which may include small companions and have low signal-to-noise. To avoid this problem, we have taken the median slope as the measure of the galaxy half-light radius.

*Apertures*

For each sub-sample we measured magnitudes, radii and then computed the average surface brightness for a range of apertures in order to estimate what is the effect on our measurements of the aperture chosen. We estimated this behavior from the trend of the mean values of magnitude, radii and SB as a function of the aperture. As the aperture exceeds the radii of the largest galaxies the above values converge. We selected the final apertures for each sample when the mean SB changed by less than 0.1 magnitudes and the mean radius by less than 5% in going from one aperture to the next largest, selecting the smaller apertures for the final values.

We define SB by this formula: $SB = m + 2.5\log(2\pi r^2)$, where r is the HLR in arc seconds measured by the slope method. This is not the same as the average SB within the HLR, but it produces a SB which can be consistently compared at all redshifts, independently of image size.

*Measuring the mean SB of the population*

Since there is considerably scatter in the SB of individual galaxies, we are using the *population* mean at each redshift as the measure of surface brightness. To estimate the population mean from the sample, we can't just use the sample mean. The samples are "censored" in that unresolved galaxies, whose number is known, are excluded from the measured SBs, because their radii are too small to be accurately resolved. There are two ways to measure the population mean from a censored sample. One is to assume that the minimum radius observable is known, and use that number to determine the population mean and variance from the sample mean and variance. There are a number of standard statistical methods to do that (for example, Cohen).

However, this calculation does depend on the assumed lowest observable radius and assumes that all the unresolved galaxies are smaller than this radius. A simpler method is to simply take the median SB of all the galaxies, including the unresolved galaxies that would otherwise be in the sample. This method merely assumes that the unresolved galaxies have SB higher than the median SB. By using the median, the method is also less sensitive to measurement errors in the smallest and largest galaxies. Like the censored analysis method, it does assume that the distribution is approximately Gaussian, so that the median is a good estimate of the mean. This is the method that we chose and in the entire following, median SB is used in this sense.

The statistical error in this estimate is simply $(\sigma/N)^{0.5}$, where N is the total of the resolved and unresolved galaxies in the sample and $\sigma$ is the standard deviation of the population. The standard deviation of the population can be estimated from that of the sample by the censored data methods. For this paper, we simply take as an estimate of the population $\sigma$ that for the sample with the largest variance. This almost certainly overestimates variance, but gives the most generous analysis of the models' fit.

*Comparison of observations with the LCDM model*

The comparison of the data with the LCDM model is complicated by the fact that in expanding-universe models of galaxy formation, it is expected that the radii of galaxies with the same absolute luminosity will grow with time of formation. Mo, Mao and White (1998) first showed that the radius of disk galaxies forming at redshift z should be a fixed fraction of the size of the dark matter halo. This in turn is proportional to $H^{-1}(z)$ for fixed viral velocity or $H^{-2/3}(z)$ for fixed mass, and somewhere in between for fixed absolute luminosity L, where

$$H(z) = H_0[\Omega_m(1 + z)^3 + \Omega_k(1 + z)^2 + \Omega_\Lambda]^{1/2} \quad (1)$$

,where $\Omega_m$ is the ratio of matter density to closure density, $\Omega_\Lambda$is the ratio of dark energy density to closure density and $\Omega_k$ is the curvature parameter, assumed to be zero for an inflationary universe. Thus the expectation for observed surface brightness becomes

$$SB= SB_0 - 2.5 \log((\Omega_m(1 + z)^3 + \Omega_k(1 + z)^2 + \Omega_\Lambda)/(1 + z)^3), \quad (2)$$

for the case of fixed virial velocity. It should be noted that in the case of of $\Omega_m = 1$, $\Omega_k = \Omega_\Lambda = 0$, eq. (2) predicts constant SB and is indistinguishable from the non-expanding prediction. However, this choice of cosmological parameters can be excluded on many other grounds, such as predicting an age of the universe of 9 Gy. In contrast, in the case of the concordance LCDM cosmology $\Omega_m$ =0.3, $\Omega_k$ =0, $\Omega_\Lambda$=0.7, the predictions differ very significantly from those of the non-expanding model. If low-z SB values are assumed to be equal for the expanding and non-expanding models, the difference between their predictions for SB at high z is 1.3 magnitudes. However, it should be noted that the if the ENE and LCDM predictions are plotted together (see Figure 2) the curves become nearly parallel at high z. This means that only a study that compares SB at high and low z can differentiate between the models.

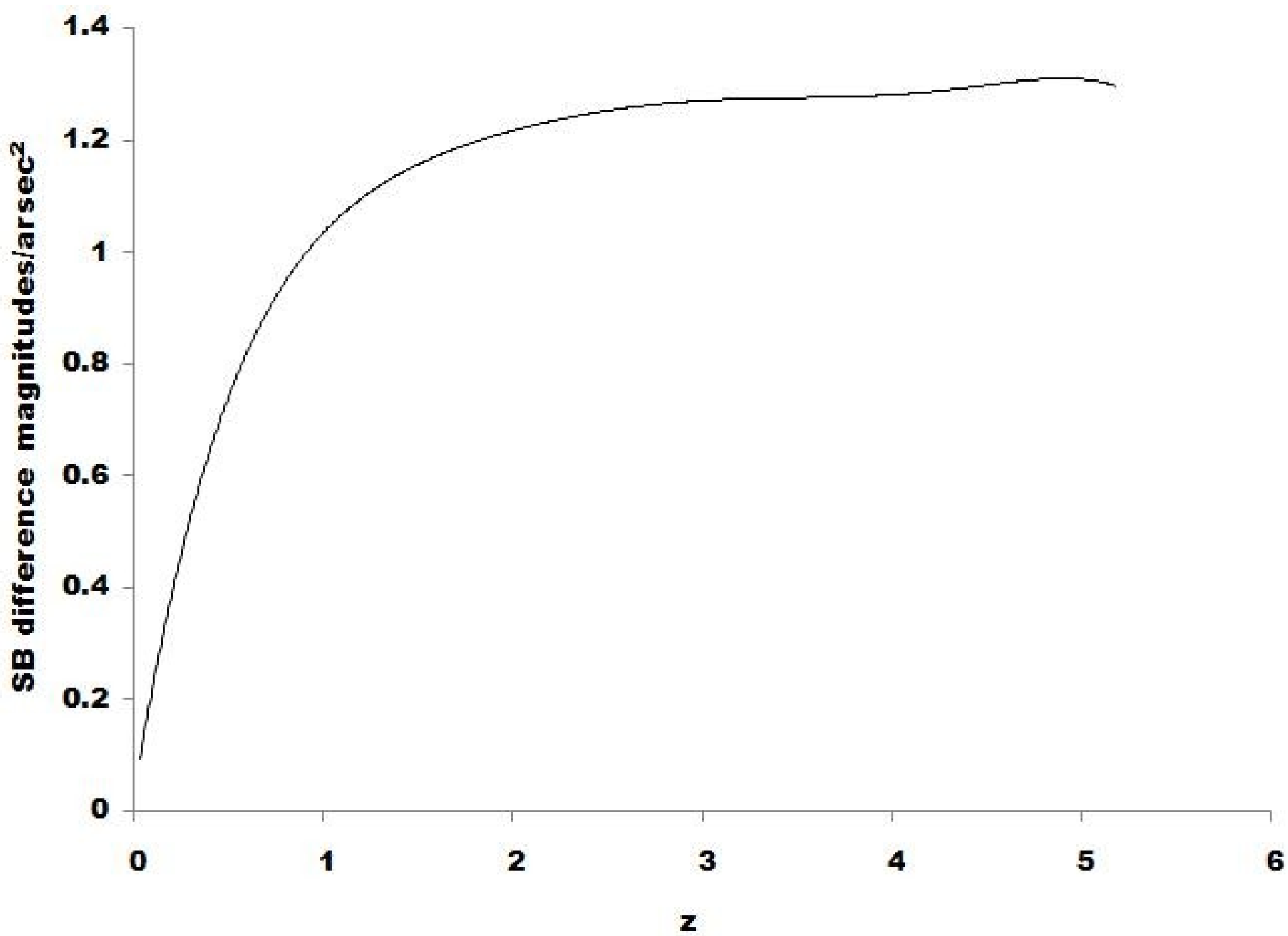

**Figure 2**. The predicted observed SB for the LCDM model with galaxy size evolution plotted against redshift. For the ENE model the prediction is a horizontal line, as SB is predicted to be constant. Note that at z>1.5, LCDM also predicts a nearly constant observed SB, so a comparison of galaxies at high and low z is essential to differentiate the two models.

We show the results for all samples in Table 1. We show here the number of unresolved galaxies, total galaxies, median SB, the predicted SB from the best fit of the LCDM model, observed minus predicted values and the $\chi^2$, using a $\sigma$ of 1.27 mag, derived from the nuv-B sample. The fit assumes that the FUV and NUV SB are the same at a given z, but fits with different FUV and NUV values are little different.

As can be seen from the $\chi^2$, the LCDM model is a very poor fit to all the data, with a formal probability of $10^{-18}$ or 9 sigma. This is clearly the result of the difference between the GALEX points and the HUDF points, as is clearly shown in Figure 3, which plots the difference between observations and the LCDM fit against z. Relative to the LCDM predictions, the points

**Table 1** Comparison of LCDM model with observation.

| sample | z | N unres | N tot | Med SB | pred | diff | $\chi^2$ |
|---|---|---|---|---|---|---|---|
| fuv 0.04 | 0.04 | 6 | 36 | 24.46 | 23.72007 | 0.739928 | 12.31 |
| fuv 0.1 | 0.1 | 13 | 38 | 24.55 | 23.65721 | 0.892791 | 18.93 |
| fuv-B | 2.03 | 5 | 39 | 20.65 | 21.36893 | -0.71893 | 12.59 |
| fuv-V | 3.13 | 3 | 54 | 19.96 | 20.41219 | -0.45219 | 6.9 |
| fuv-i | 4.34 | 3 | 26 | 19.35 | 19.59413 | -0.24413 | 0.96 |
| fuv-z | 5.18 | 6 | 21 | 18.94 | 19.1241 | -0.1841 | 0.44 |
| | | | | | | | |
| nuv 0.1 | 0.1 | 15 | 75 | 24.53 | 23.65721 | 0.872791 | 35.7 |
| nuv-B | 1.05 | 1 | 23 | 22.31 | 22.46881 | -0.15881 | 0.36 |
| nuv-V | 1.83 | 1 | 26 | 21.19 | 21.57241 | -0.38241 | 2.37 |
| nuv-i | 2.77 | 3 | 52 | 20 | 20.69836 | -0.69836 | 15.85 |
| nuv-z | 3.32 | 2 | 41 | 19.91 | 20.27009 | -0.36009 | 3.32 |

at low z are about 1.2 magnitudes/arsec$^2$ too dim or, to put it another way, the galaxies are 1.7 times too large, as compared with the high-z samples. If only the HUDF points are fit, the $\chi^2$ drops to only 6.88, clearly a good fit, and the predicted low-z SB is 23.32 magnitudes/arsec$^2$, instead of the observed 24.5 magnitudes/arsec$^2$

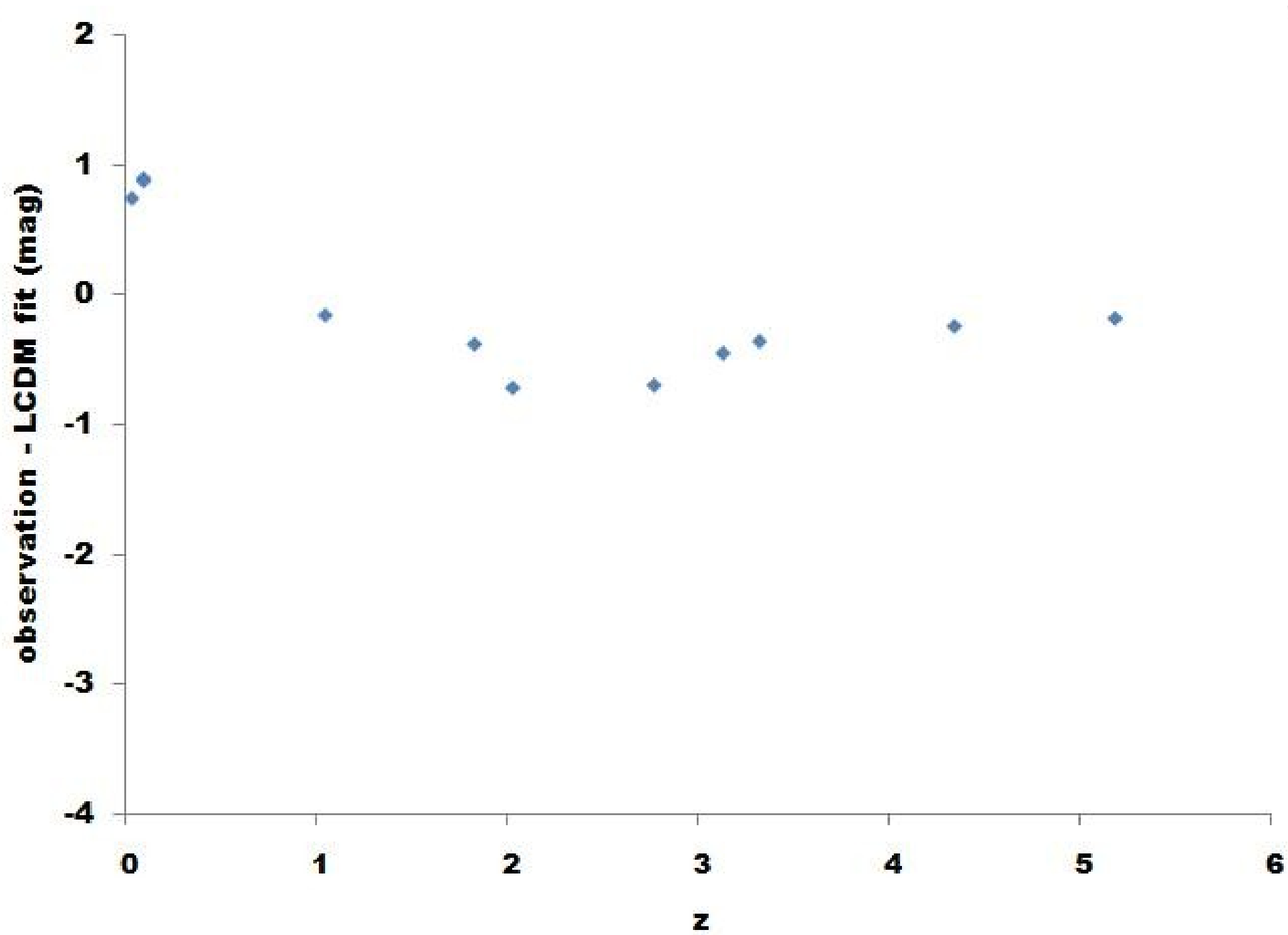

**Figure. 3** Summary of surface brightness comparison with LCDM model. The difference between observed SB and best LCDM fit in magnitudes/arsec$^2$ is plotted against redshift.

This prediction was derived using eq. 6, based on $H^{-1}(z)$ size evolution for fixed virial velocity. However substituting a $H^{-2/3}(z)$ size evolution for fixed mass or anything in between these two formulas would make the predicted size evolution smaller and therefore would make still worse the disagreement between prediction and data.

*Comparison of observations with ENE model*

We next compare the data with the ENE model. Table 2 summarizes the comparison samples and the measurement of their median surface brightness. Here, the comparison is simpler, since the ENE prediction is a constant SB. The columns give unresolved galaxies, total number of galaxies, median observed SB, and the $\chi^2$, determined based on the best fit to a constant SB, which is 24.75. Figure 4 plots observed median SB against z.

For the data as a whole, the ENE model is a significantly better fit than the LCDM model. The ratio of $\chi^2$ for LCDM to that for ENE is 4.85, which for 10 degrees of freedom in each set has a formal probability of existing by chance of 1%. However, for the data as a whole, the ENE model is also not an acceptable fit, since the $\chi^2$of 22.60 also has only a 1% chance of resulting from a constant SB. The unacceptability of the fit is entirely due to the two highest-z points. For the z range from 0.03 to 3.5 the $\chi^2$ of 12.84 is an acceptable fit for 8 degrees of freedom.

**Table 2** Comparison of ENE model with SB observations

| Sample | z | N unres | N total | med SB | $\chi^2$ |
|---|---|---|---|---|---|
| fuv 0.04 | 0.04 | 6 | 35 | 24.58 | 0.70875 |
| fuv 0.1 | 0.1 | 13 | 53 | 25.04 | 2.597 |
| fuv-B | 2.03 | 5 | 44 | 24.51 | 1.71875 |
| fuv-V | 3.13 | 3 | 91 | 24.58 | 1.84275 |
| fuv-i | 4.34 | 3 | 67 | 25.12 | 5.427 |
| fuv-z | 5.18 | 6 | 17 | 25.25 | 2.551062 |
| | | | | | |
| nuv 0.1 | 0.1 | 15 | 85 | 24.89 | 0.897812 |
| nuv-B | 1.05 | 1 | 28 | 24.48 | 1.372 |
| nuv-V | 1.83 | 1 | 37 | 24.51 | 1.445313 |
| nuv-i | 2.77 | 3 | 83 | 24.5 | 3.50675 |
| nuv-z | 3.32 | 2 | 70 | 24.87 | 0.529375 |

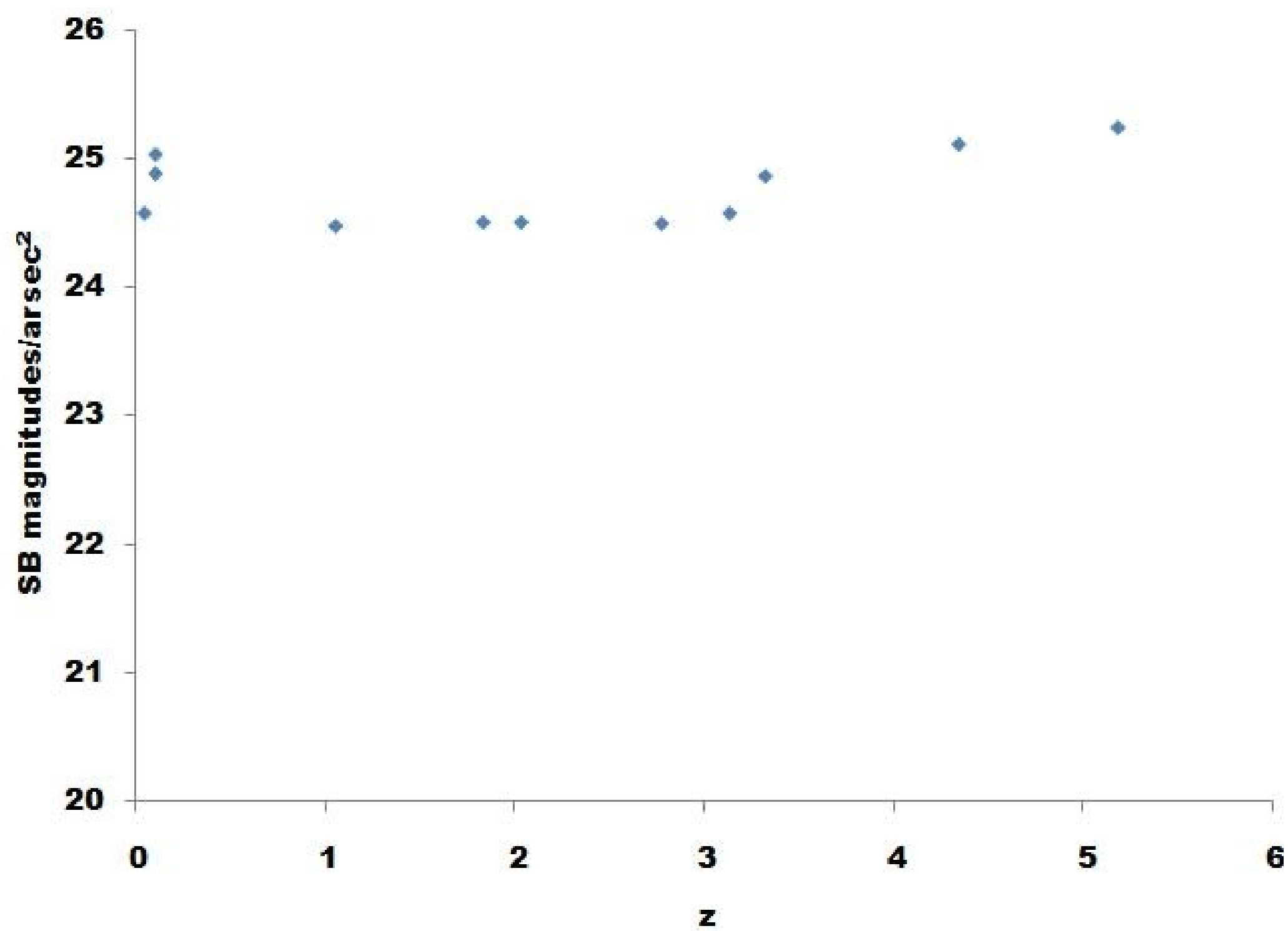

**Figure 4** Observed median SB against redshift. The ENE model predicts a constant SB.

It is also important to note that, using the ENE analysis, there is a statistically significant difference between the SB of the GALEX samples at z=0.1 and that at z=0.04. Indeed, the difference between the low and high z samples from GALEX is almost the same as the difference between the bulk of the HUDF samples and the two high-z points. Since in both the cases of z=0.1 and z=4-5 the median radius of the sample is relatively close to the resolution of the telescope (less than a factor of 2 bigger) while in all the other cases, there is a greater gap between the median and smallest observable radius, we suspected that the 0.5 magnitude difference could be due to overestimation of the radii of galaxies with relatively small image sizes. This could be due to, for example, the inclusion in the small-image-size samples of galaxies that are actually multiple and which would be resolved and excluded from the sample if the images were larger relative to the resolution or the pixel size. We have not been able to confirm this hypothesis as yet.

However, if all the low-resolution points are removed (those with a small gap between the telescope resolution and the median observed radius) there is a remarkable constancy of SB, with a $\chi^2$of 4.39 for 6 degrees of freedom.

## 4. Conclusions and open questions

The facts that the predictions of the LCDM model are so far from fitting the data and that the ENE model is much closer to the data are not at all the results that one would expect from the widespread acceptance of the LCDM model. The results point in the same direction as the earlier result of Lerner, which is to pose a severe challenge, at the least, to the LCDM model and, in fact to any model that explains the Hubble relation from the expansion of the universe and points to the possible viability of the ENE model. However, this conclusion must be seen as preliminary and tentative, and some open questions must certainly be answered before definitive conclusion scan be drawn on the Tolman test of the two models.

In the case of the LCDM model, the failure to fit the data is due entirely to the difference between the GALEX and HUDF SB measurements. Since there is no data in the data sets we have used between z=0.1 and z=0.9, it is reasonable to ask : what happens in this gap? If there are datasets that cover all or most of this redshift region, in which the predictions of the two models are widely divergent, this could be an additional convincing test of the LCDM model.

In fact the GEMS data set could provide such an additional test. The GEMS data set covers a region 75 times large in area than HUDF, so it allows significant sample sizes for very bright galaxies at much lower redshift than HUDF. Two images were taken, one in the z –band and the other in the V-band. It should be possible to use these images to compare the SB in the rest-frame B-band at z=0.9 with that at z=0.4, thus filling in most of the gap between HUDF and GALEX. (Lopez-Corriodora has used this same data set for a test of the angular-radius-redshift realtionship). We intend to do this.

Second, for the ENE model, there remains the question of whether the small images are mismeasured, a result which would also have some effect on the LCDM analysis. A more through analysis of the change in SB with z in the GALEX sample should shed light on this question, and will be pursued in the future.

Acknowledgement. The author wishes to acknowledge the large contributions to the analysis of this data of Renato Falomo and Riccardo Scarpa.